\documentclass[12pt,preprint]{aastex}

\def\vs{$v_{s}$\,\,}

\def\dem71{DEM L 71\,}

\def\tetp{($T_{e}/T_{p}$)$_{0}$\,}

\begin{document}

\title{The Heating of Thermal Electrons in Fast Collisionless Shocks: The
Integral Role of Cosmic Rays}


\author{Cara E. Rakowski\altaffilmark{1}}
\affil{Space Science Division, Naval Research Laboratory Code 7674R,
Washington, D.C. 20375} 
\author{J. Martin Laming}
\affil{Space Science Division, Naval Research Laboratory Code 7674L,
Washington, D.C. 20375} 

\and

\author{Parviz Ghavamian}
\affil{Department of Physics and Astronomy, Johns Hopkins University,
3400 N. Charles Street, Baltimore, MD, 21218}
\email{parviz@pha.jhu.edu}

\altaffiltext{1}{National Research Council Fellow}

\begin{abstract}
Understanding the heating of electrons to quasi-thermal energies at
collisionless shocks
has broad implications for plasma astrophysics.
It directly impacts the interpretation of X-ray spectra from shocks,
is important for understanding how energy is partitioned between the
thermal and cosmic ray populations, and provides insight into the
structure of the shock itself.
In \citet{GLR07}
we presented observational evidence for an inverse square
relation between the electron-to-proton temperature ratio and the
shock speed at the outer blast waves of supernova remnants
in partially neutral interstellar gas. There we outlined how
lower hybrid waves generated in the cosmic ray precursor could
produce such a relationship by heating the electrons to a common
temperature independent of both shock speed and the strength of the
ambient magnetic field. Here we explore the mechanism of
lower hybrid wave heating of electrons in more detail. Specifically we examine
the growth rate of the lower
hybrid waves for both the kinetic (resonant) and reactive cases.
We find that only the kinetic case
exhibits a growing mode.   At low Alfv\'{e}n Mach numbers ($\sim$15)
the growth of lower hybrid waves can be faster than the
magnetic field
amplification by modified Alfv\'{e}n waves.

\end{abstract}


\section{Introduction}

The
main accelerators of
cosmic rays (CRs) are widely
believed to be high Mach number shocks in collisionless plasma, here
loosely defined as plasma where charged particles interact
predominantly through plasma waves rather than by Coulomb collisions
\citep[for a thorough conceptual and historical review
see][]{MalkovDrury2001}. However, a concensus is emerging that
CRs are not simply a byproduct of collisionless shocks, but
in fact
play an integral role in the shock structure, dynamics, and
energetics. For example sound waves in a cosmic ray pressure
gradient can smooth out the shock jump in cosmic ray modified shocks
\citep{Drury86}. More recent analytic and
numerical work has shown that modified Alfv\'{e}n waves in the CR precursor may
amplify the magnetic field to many times its ambient value by
generating perpendicular magnetic field from an initially
quasi-parallel geometry 
\citet{LucekBell2000,BellLucek2001,Bell2004,Bell2005}.
Observational support for dramatic magnetic field amplification ahead of shocks
exists in the form of extremely thin X-ray synchrotron rims of
supernova remnants (SNRs) such as Cassiopeia A \citep{VinkLaming2003}, SN
1006 \citep{Long2003, Yamazaki2004}, and Tycho's SNR (Warren et al. 2005; Cassam-Chena{\"i} et al. 2007).
As noted by \citet{Cassam2007} there are two possible interpretations for the narrow width,
however both require a dramatically amplified magnetic field ahead of the shock. Either the rims are thin
because the high magnetic field causes rapid synchrotron
cooling of the X-ray emitting electrons or the scale of the rims represents
the scale of magnetic field de-amplification behind the shock.

In this paper we explore another area where CRs may influence the
properties of the shock, namely through the heating of quasi-thermal
electrons.
For shocks in collisionless plasma the
heating of electrons must occur through the damping of waves
generated by the other more massive charged particles that dominate
the energetics. Given the wide array of possible plasma
instabilities at collisionless shocks an observational relationship
for electron temperature, $T_{e}$ at the shock front was required to
limit theoretical discussions. An inverse relationship between the
initial ratio of electron to proton temperatures immediately behind
the shock,
\tetp,
and the shock velocity,
\vs,
 has been reported in a series of
observational papers on supernova remnant
shocks \citep{G01,G02,GRHW03,RGH03,R05,GLR07} and has also been noted
among the higher Alfv\'en Mach number events in a sample of solar
wind shocks \citep{Schwartz88}. In \citet{GLR07}, we focused on
shocks propagating into partially neutral gas.
Here the collisional excitation of broad and narrow Balmer line emission at the shock
front can be used to diagnose
\tetp.  In \citet{G01}, \citet{G02}
and \citet{GLR07} we described the method of simultaneously
constraining \vs\, and \tetp\, via measurement of the
width of the broad Balmer line and the ratio of broad to narrow
Balmer line flux \citep[see also][]{Heng07a,Heng07b}.
Our results are consistent with an inverse square
relationship, \tetp $\propto 1/v_{s}^{2}$ for shock speeds above
$\sim$400~km~s$^{-1}$ (Ghavamian, Laming \& Rakowski 2007). Given
that $T_{p} \propto v_{s}^{2}$ at the shock front by the
Rankine-Hugoniot jump conditions, the inverse relationship between
equilibration and shock speed implies that the electron temperature
itself is nearly {\it constant} $\sim$0.3 keV, independent of shock
speed.

The insensitivity of electron temperature to shock velocity suggests a heating mechanism
within the extended diffusive cosmic ray precursor ahead of the shock.
In this case the
electron heating would be more reflective of the generic properties of cosmic ray
acceleration and diffusion than tied to the specific attributes of the shock.
In contrast, prior work on
heating by shock-reflected ions that are confined to within a gyroradius of
the (quasi-perpendicular) shock \citep{Cargill88}
suggested that \tetp\, would remain constant
with shock velocity.
In \citet{GLR07} we suggested that lower hybrid waves in the cosmic ray
precursor of a perpendicular shock might be a plausible electron heating
mechanism.

Lower hybrid waves are electrostatic ion waves directed
nearly perpendicular to the magnetic field with a frequency equal
to the geometric mean of the electron and ion gyrofrequencies.
Electrons that would
otherwise screen the ion oscillation are pinned to magnetic field lines.
In addition, the group velocity parallel to the magnetic field greatly exceeds the group
velocity perpendicular to the field ($\omega /k_{\|} >>\omega /k_{\perp}$).
Therefore, the wave can simultaneously resonate with ions moving across the
field lines and electrons moving along the field lines, facilitating collisionless
energy exchange between them. Based on simple arguments about the width of the
cosmic ray precursor and the electron diffusion along the field
lines, we showed that electron heating from lower hybrid waves in
the cosmic ray precursor would be independent of both the shock
speed and the magnetic field. Here we explore this mechanism in
more physical detail.

In Section 2 we calculate the growth rate of lower hybrid
waves, first examining the kinetic (resonant) then the reactive
(non-resonant) case.
The treatment here is mathematically similar to the work
on modified Alfv\'{e}n waves by \citet{Achterberg83} and
\citet{Bell2004,Bell2005} involving the cosmic ray contribution to the plasma dielectric
tensor.
 The analysis also draws on the work of \citet{L01a}
and \citet{L01b} on lower hybrid waves from shock-reflected ions.
We compare these growth rates with those for magnetic field amplification,
to assess the conditions under which electron heating might occur.
In section 3 we discuss the structure of the cosmic ray shock precursor
in more detail. We pay particular attention to the magnetic field geometry,
since the excitation of lower  hybrid waves requires a
quasi-perpendicular shock. We show schematically how magnetic field
amplification and lower-hybrid wave heating might co-exist in the shock
precursor for either parallel or perpendicular initial geometries.
We also review some other ideas for electron heating, and make some
quantitative predictions from our model for various shock
parameters. Included in the appendices are a
discussion of cosmic ray diffusion coefficients and a derivation of
the resonant growth rate for electromagnetic waves.

\section{Cosmic Ray Growth Rate of Lower Hybrid Waves}
Lower hybrid waves ahead of collisionless shocks have particularly interesting properties.
They can have a group velocity away from the shock equal to the
shock velocity itself \citep{mcclements97}. {\it This can in principle allow the waves to grow to
large amplitudes, even if their intrinsic growth rate is small.}
To determine if lower hybrid waves can heat the electrons to the
$\sim$0.3 keV temperature observed, we must first calculate the
growth rate of this instability to see if it will have sufficient
power to overcome the damping effect of the electrons as well as
to compete with other instabilities in the precursor.
We calculate this growth rate in both kinetic
and reactive limits, i.e. either considering the
cosmic rays with energies in resonance with the lower hybrid wave frequency or the
integrated contribution of the entire distribution, respectively
\citep[see e.g][]{melrose86}.
Related kinetic and reactive cases were calculated in \citet{L01a}
and \citet{L01b},  but only for the case of shock reflected,
non-relativistic ions
gyrating around the magnetic field, represented as a particle beam.
Here we begin the discussion with the resonant case.

\subsection{Kinetic Growth Rate}

We model the normalized cosmic ray distribution function diffusing upstream as

\begin{equation}
f(p) = \frac{n_{CR}^{\prime}}{4\sqrt{2}(\pi \kappa)^{3/2}p_{t}^{3}}
\frac{(2\kappa-3)\Gamma(\kappa)}{\Gamma(\kappa -1/2)}
\left[ 1 + \frac{(p_{x}-mv_{s})^{2} + p_{y}^{2} +p_{z}^{2}}
                  {2\kappa p_{t}^{2}} \right]^{-\kappa}
\end{equation}
where the coordinate system is aligned so that the shock speed $v_s$ lies in the
$x$-direction.  Here $p_t$ is defined as the ``thermal'' momentum, which we take
to be $\left(3/4\right) mv_s$, and $n_{CR}^{\prime}$ denotes the density of
suprathermal particles with distribution function $f\left(p\right)$.
The functional form above, known as a
``kappa'' distribution, is often seen for particle distribution
functions associated with shocks in the solar wind, and may be derived as equilibrium
distributions for a system of particles and waves under certain conditions
\citep[see e.g.][]{laming07}, in contrast to
a system of particles only which gives a Maxwellian distribution
of width $p_t$.
The kappa
distribution resembles a Maxwellian for $p < p_t$, and in fact for $\kappa\rightarrow
\infty$ is exactly a Maxwellian. At higher particle momenta it tends
smoothly to a distribution $f\propto p^{-2\kappa }$. Below we shall take $\kappa =2$
to model the well known $f(p)\propto p^{-4}$ cosmic ray distribution predicted by
diffusive shock acceleration in shocks with a compression ratio of 4.
In connecting the cosmic rays to the
lower energy particles in this way, we are somewhat blurring the
distinction between ``cosmic rays'' and other suprathermal particles reflected from
the shock. Hence we denote the combined density of these particles as $n_{CR}^{\prime}$
to distinguish it from density of true cosmic rays, $n_{CR}$, which will appear
in expressions derived by other authors.
Note that all these particles
are distinct from the ambient thermal plasma upstream of the shock, which here is
considered to be a Maxwellian with much lower temperature than $p_t$ used in
Equation (1) for the upstream suprathermals.
When discussing the kinetic instability we will be
focusing on particles that obey a diffusion equation ahead of the shock, due to
their interaction with turbulence, rather than gyrating around field
lines.  We will qualify this distinction more carefully
below in our discussion of the reactive instability.

The appropriate dispersion relation for Equation (1) can be found from
the cold plasma dieletric tensor. For electrostatic waves at frequencies
close to the lower hybrid wave frequency we have \citep{L01a,L01b}
\begin{equation}
K_{L} = 1 + \frac{\omega_{pe}^{2}}{\Omega_{e}^{2}} \sin^{2}\theta
 - \frac{\omega_{pi}^{2}}{\omega^{2}}
 - \frac{\omega_{pe}^{2}}{\omega^{2}}\cos^{2}\theta
+ \frac{4\pi q^{2}}{k^{2}} \int \frac{{\bf k} \cdot \partial f/\partial \bf p}
    {\omega - \bf k  \cdot \bf v } d^{3}{\bf p}  = 0
\end{equation}
where $\theta$ denotes the angle between the wavenumber
of the perturbation
and the preshock magnetic field;
$\omega_{px} = (4\pi q^2 n_{x}/m_{x})^{(1/2)}$ is the plasma frequency of a
given species $x$
(electrons, ions, cosmic rays...) with charge $q$, density $n_{x}$, and mass $m_{x}$;
$\Omega_{x} = qB/(m\gamma_L c)$ is the cyclotron frequency of species $x$
(with $\gamma_L$ being the Lorentz factor);
and unadorned $\omega$ being the lower hybrid wave frequency which is the
geometric mean of the electron and proton cyclotron frequencies.
Using the Landau prescription for evaluating the integral at the resonant pole and taking
only the imaginary parts of the dielectric tensor equation, we find
the growth rate for the lower hybrid waves
\begin{equation}
\gamma = \frac{2\pi^{2}q^{2}}{k^{2}} \frac{\omega^{2}}{\omega_{pi}^{2}+
\omega_{pe}^{2}\cos^{2}\theta} \int \delta(\omega - {\bf k} \cdot {\bf v})
  {\bf k} \cdot \frac{\partial f}{\partial {\bf p}} d^{3}{\bf p}.
\end{equation}

To compute $\gamma$, we take $n_{CR}^{\prime}\propto\exp{-xv_s/D}$ where $D$
is the cosmic ray diffusion coefficient and is in principle
dependent on the cosmic ray momentum  (making $l=D/v_{s}$ the characteristic
diffusive lengthscale).
With this substitution we begin the evaluation of the last integral in Equation (3)
\begin{equation}
\int^{\infty}_{0} 2\pi p_{\bot}fdp_{\bot} =
\frac{n_{CR}^{\prime}}{4\sqrt{2}(\pi \kappa)^{3/2}p_{t}^{3}}
\frac{(2\kappa-3)\Gamma(\kappa)}{\Gamma(\kappa -1/2)} \frac{2\pi
\kappa}{\kappa -1} p_{t}^{2} \left[ 1 +
\frac{(p_{x}-mv_{s})^{2}}{2\kappa p_{t}^{2}} \right]^{1-\kappa}
e^{-xv_{s}/D},
\end{equation}
where we have
separated
out the components of ${\bf p}$ perpendicular to the shock
(and ${\bf k}$) from $p_x$.
Substituting back into Equation (3) yields a growth rate
\begin{eqnarray}
& & \gamma = \left( \frac{\pi}{\kappa}\right)^{3/2}\frac{q^{2}}{p_{t}k}
\frac{\omega^{3} n_{CR}^{\prime}}{\omega_{pi}^{2}+ \omega_{pe}^{2}\cos^{2}\theta}
\frac{(2\kappa-3)\Gamma(\kappa)}{\sqrt{2}\Gamma(\kappa -1/2)}
\int \delta(\omega - {\bf k} \cdot {\bf v}) \nonumber \\
& & \times \left\{ -\left[1+\frac{(p_{x}-mv_{s})^{2}}{2\kappa
p_{t}^{2}}\right]^{-\kappa} \frac{(p_{x}-mv_{s})}{p_{t}^{2}} +
\frac{\kappa}{\kappa -1}\frac{xv_{s}}{D^{2}}\frac{\partial
D}{\partial p_{x}} \left[ 1 + \frac{(p_{x}-mv_{s})^{2}}{2\kappa
p_{t}^{2}} \right]^{1-\kappa} \right\} e^{-xv_{s}/D} dp_{x}.
\end{eqnarray}
For waves to stay in contact with the shock
$w/k\simeq -2v_{s}$ \citep{L01a} in the cold plasma electrostatic limit, i.e. reflected ions returning to the shock
excite the waves. This remains generally true when
these approximations are
relaxed \citep{L01b}, so the $\delta$ function picks out $p_{x} = - 2mv_{s}$
(cosmic rays returning to the shock). Hence
\begin{eqnarray}
& & \gamma = \left( \frac{\pi}{\kappa}\right)^{3/2}
\frac{q^{2}\omega^{3} n_{CR}^{\prime} v_{s}^{2}m_{CR}}
{p_{t}(\omega_{pi}^{2}+ \omega_{pe}^{2}\cos^{2}\theta)}
\frac{(2\kappa-3)\Gamma(\kappa)}{\sqrt{2}\Gamma(\kappa -1/2)}
\nonumber \\
& & \times\left[1+ \frac{9m^{2}v_{s}^{2}}{2\kappa p_{t}^{2}} \right]^{-\kappa}
\left\{ \frac{3mv_{s}}{p_{t}^{2}} + \frac{\kappa xv_{s}}{(\kappa-1) D^{2}}
\frac{\partial D}{\partial p_{x}}
\left( 1+ \frac{9m^{2}v_{s}^{2}}{2\kappa p_{t}^{2}} \right)
\right\} e^{-xv_{s}/D}
\end{eqnarray}
Assuming $\partial D/\partial p_{x} = 0$ (see Appendix A
on cosmic ray diffusion coefficients) and $\kappa = 2$, only the
first term within curly brackets remains. Evaluating the $\Gamma$ functions
and substituting $p_{t} = (3/4)mv_{s}$ we arrive at
the following expression for the kinetic growth rate of lower hybrid waves
\begin{equation}
\gamma =  \frac{8}{225} \frac{\omega_{pi}^{\prime 2} \omega}{\omega_{pi}^{2}
+ \omega_{pe}^{2}\cos^{2}\theta} e^{-xv_s{}/D}.
\end{equation}
where $\omega_{pi}^{\prime}$ denotes the plasma frequency for
$n_{CR}^{\prime}$. Substituting in the frequency definitions we note
that approximately $\gamma \propto (n_{CR}^{\prime}/ n_{i})\Omega_{i}$.
Before proceeding, we pause to compare this growth rate with those for
magnetic field amplification. In the case where Alfv\'en waves are resonantly
excited, the growth rate is \citep{melrose86,Pelletier2006}
\begin{equation}
\gamma _{B,res}={3\pi\over 16}{\Omega _i\over v_A}{n_{CR}\over n_i}
\left({\cos\theta\over |\cos\theta |}v_s\cos\phi -{4\over 3}v_A-
{\pi\over 4}v_s\sin\phi\right)k_{||}r_g
\end{equation}
where $v_A$ is the Alfv\'en speed, $n_i$ is the density of ions in the background
plasma and $r_g$ is the gyroradius of cosmic rays. This expression differs from that
in the cited references in the factor $\cos\phi$ and the term in $\sin\phi$,
where $\phi$ is the angle between
the shock velocity and the magnetic field. At perpendicular shocks,
the growth rate of resonant Alfv\'en waves
can be neglected, but at parallel shocks may be larger than that for lower hybrid
waves, depending on the ratio $n_{CR}^{\prime}/n_{CR}$.  However as we shall argue
later, all shocks subject to magnetic field amplification become perpendicular, and this
is the geometry where lower hybrid waves are most effectively excited, so we neglect
$\gamma _{B,res}$ from here onward. \citet{Bell2004} discovered a nonresonant growth rate
for Alfv\'en waves, with approximate growth rate
\begin{equation}
\gamma _{B,nonres}=\sqrt{{n_{CR}\over n_i}k_{||}v_s\Omega _i-k_{||}^2v_A ^2}
\end{equation}
which has a maximum value of $M_A\Omega _i n_{CR}/2n_i$.
According to \citet{Bell2005}, this instability operates for arbitrary orientations
of ${\bf B}$, ${\bf v_s}$ and ${\bf k}$, indicating that it will also amplify
magnetic field at perpendicular shocks.
Its growth rate is
strongest for ${\bf k}||{\bf B}$
and zero for ${\bf k}\perp {\bf B}$. Equating the maximum value of $\gamma _{B,nonres}$
with the lower hybrid wave growth rate calculated above, we find
the critical Alfv\'enic Mach number
$M_A\simeq 3n_{CR}^{\prime}/n_{CR}$, such that for higher $M_A$, cosmic rays
preferentially amplify magnetic field, and for lower $M_A$ they generate lower hybrid
waves. The numerical value depends on the ratio $n_{CR}^{\prime}/n_{CR}$. In the
next section we will argue that these two densities should not be the same, and that
$n_{CR}^{\prime}>n_{CR}$, following from a consideration of the reactive growth rate
for lower hybrid waves.

\subsection{Reactive Growth Rate}

The reactive case involves the integrated contribution to the growth rate from the entire cosmic ray distribution.
Thus we will examine successive orders in an expansion of $f(p)$ to see if they produce any growing modes. Although no
instability is found in this process, we do uncover potentially interesting constraints on the properties of $f(p)$.

We consider again the last term in Equation (2), the cosmic ray contribution to the dielectric tensor
\citep[e.g.][]{melrose86} which includes a factor that reduced to unity for the resonant case;
\begin{equation}
K_{L}^{CR} = \frac{4\pi q^{2}}{k^{2}}
    \int \frac{{\bf k} \cdot \partial f/\partial \bf p}
  {\omega - \bf k  \cdot \bf v } \frac{\bf k  \cdot \bf v}{\omega} d^{3}{\bf p}.
\end{equation}
For the case of a beam of cosmic ray particles localized around $\bf v_{s}$, one
recovers the usual beam reactive instability \citep[equation (A6)]{L01b}. However, as
we demonstrate below, for
a more physical quasi-isotropic cosmic ray distribution no instability is recovered.
We consider the case where cosmic rays drift with velocity ${\bf v_s}$, and
waves are generated with ${\bf k}\| {\bf v_s}$.
The CR distribution function from appendix A, expanded in terms of the cosine
of the angle between ${\bf k}$ and ${\bf v}$ or ${\bf k}$ and ${\bf v_{s}}$, $\cos\alpha = \mu$, is
\begin{equation}
f = f_{0} + \mu \frac{\partial f}{\partial \mu} + ...
= f_{0}(1 + \mu \frac{3v_{s}}{v} + ...),
\end{equation}
such that $\int \mu v f d \Omega = f_{0}v_{s}$ as before.
Neglecting terms of order $v_{s}^2/v^{2}$, the CR contribution to the dielectric
tensor becomes
\begin{equation}
K_{L}^{CR} \approx \frac{4\pi q^{2}}{k^{2}} \int_{0}^{\infty} 2\pi
    \int_{-1}^{1} \frac{vk^2}{\omega}
 \frac{(\mu^{2}+3\mu^{3}v_{s}/v)}{\omega - k v \mu } d\mu
\frac{\partial f_{0}}{\partial p}p^{2}dp.
\end{equation}
Expanding the numerator into terms divisible by $(\omega/kv - \mu)$ and
evaluating the integral over $\mu$ we obtain
\begin{equation}
K_{L}^{CR} \approx \frac{8\pi^{2} q^{2}}{\omega} \int_{0}^{\infty}
\left\{\frac{-2v_{s}}{kv} - \left(\frac{3v_{s}\omega}{kv}+v\right)\frac{2\omega}{k^{2}v^{2}}
-\left(\frac{3v_{s}\omega}{kv}+v\right)\frac{\omega^{2}}{k^{3}v^{3}}
\ln \left\vert \frac{\omega-kv}{\omega+kv}\right\vert
\right\} \frac{\partial f_{0}}{\partial p}p^{2}dp.
\end{equation}
We then evaluate this integral in the two limiting cases away from the pole,
$\omega\gg kv$ and $\omega\ll kv$.
For $\omega\gg kv$;
\begin{equation}
\ln \left\vert \frac{\omega-kv}{\omega+kv}\right\vert
\simeq \frac{-2kv}{\omega} - \frac{2}{3}\left(\frac{kv}{\omega}\right)^{3}- ...
\end{equation}
leading to
\begin{equation}
K_{L}^{CR} \approx \frac{8\pi^{2} q^{2}}{\omega} \int_{0}^{\infty}
\left\{\frac{-2v_{s}}{kv} - \left(\frac{3v_{s}\omega}{kv}+v\right)\frac{2\omega}{k^{2}v^{2}}
+\left(\frac{3v_{s}\omega}{kv}+v\right)
\left( \frac{2\omega}{k^{2}v^{2}}+ \frac{2}{3\omega}\right)
\right\} \frac{\partial f_{0}}{\partial p}p^{2}dp .
\end{equation}
All terms in brackets cancel save for one, giving
\begin{equation}
K_{L}^{CR} \approx \frac{8\pi^{2} q^{2}}{\omega} \int_{0}^{\infty}
\frac{2v}{3\omega}
\frac{\partial f_{0}}{\partial p}p^{2}dp\,\,=\,\,\frac{16\pi^{2} q^{2}}{3 \omega^{2}}
\left\{ \left[vp^{2}f_{0}\right]_0^{\infty}
-\int_{0}^{\infty} 3f_{0}p^{2}\frac{dp}{\gamma m}
\right\}
\end{equation}
%
the first term goes to zero
so long as $f_{0}(\infty)\rightarrow 0$ faster than $p^{-3}$,
and the second term is $\propto n_{CR}/(\gamma m)$ leaving
\begin{equation}
K_{L}^{CR} \approx - \omega_{pCR}^{2}/\omega^{2}.
\end{equation}
and
\begin{equation}
\omega ^2\left(1+{\omega _{pe}^2\over\Omega _e^2}\sin ^2\theta\right)-\omega _{pi}^2
-\omega _{pe}^2\cos ^2\theta -\omega _{pCR}^2=0.
\end{equation}
This simply modifies the $1/\omega^{2}$ term in the dielectric tensor,
changing the frequency of the solution but not creating any complex roots,
hence no instability is generated.

Likewise, in the case where $\omega \ll kv$;
\begin{equation}
\ln \left\vert \frac{\omega-kv}{\omega+kv}\right\vert
\simeq -\frac{2\omega}{kv} -...
\end{equation}
leading to
\begin{equation}
K_{L}^{CR} \approx \frac{8\pi^{2} q^{2}}{\omega} \int_{0}^{\infty}
\left\{\frac{-2v_{s}}{kv} - \left(\frac{3v_{s}\omega}{kv}+v\right)\frac{2\omega}{k^{2}v^{2}}
+\left(\frac{3v_{s}\omega}{kv}+v\right)
\frac{2\omega^{3}}{k^{4}v^{4}}+...
\right\} \frac{\partial f_{0}}{\partial p}p^{2}dp .
\end{equation}
All but the first term are negligible in this limit, hence
\begin{equation}
K_{L}^{CR} \approx \frac{16\pi^{2} q^{2}v_{s}}{\omega k} \int_{0}^{\infty}
\frac{1}{v} \frac{\partial f_{0}}{\partial p}p^{2}dp .
\end{equation}
This leads to a full dispersion relation that can be written as
\begin{equation}
\omega^{2}\left(1+\frac{\omega_{pe}^{2}}{\Omega_e^{2}}\sin^{2}\theta\right) -
\omega \frac{4\pi q^{2}v_{s}}{k} 4\pi \int_{0}^{\infty}
\frac{1}{v} \frac{\partial f_{0}}{\partial p}p^{2}dp
-\left(\omega_{pi}^{2}+ \omega_{pe}^{2}\cos^{2}\theta \right) = 0
\end{equation}
which also lacks complex roots, regardless of the actual evaluation
of the integral.

Higher order terms in the expansion of $f=f_0\left(1+\mu\left(3v_s/v\right)
+\mu ^2\left(3v_s/v\right)^2/2+\mu^3\left(3v_s/v\right)^3/6+ ...\right)$ give rise to
higher order terms in
$\omega$. For $\omega >> kv$, the dispersion relation equation 18 becomes to next
highest order
\begin{eqnarray}
\omega ^3\left(1+{\omega _{pe}^2\over\Omega _e^2}\sin ^2\theta\right)&-\left(\omega _{pi}^2
+\omega _{pe}^2\cos ^2\theta +\omega _{pCR}^2-{72\over 5}\pi ^2q^2v_s^2\int{1\over v}
{\partial f_{0}\over\partial p}p^2dp\right)\omega\cr &+
{72\over 7}\pi ^2q^2kv_s^3\int{1\over v}
{\partial f_{0}\over\partial p}p^2dp=0,
\end{eqnarray}
which is stable since the terms $\propto \int{1\over v}
{\partial f_{CR}\over\partial p}p^2dp$ are of order $\sim \omega _{pCR}^2v_s^2/v^2
<< \omega _{pCR}^2$. When $\omega << kv$ the dispersion relation Equation (22) takes
on the next highest order terms
\begin{equation}
-{24\pi ^2q^2v_s^2\over k^4}\int {1\over v^5}{\partial f_0\over\partial p}p^2dp
\left(3\omega ^4 +\omega ^3kv_s\right) -{24\pi ^2q^2v_s^2\over k^2}
\int {1\over v^3}{\partial f_0\over\partial p}p^2dp
\left(\omega ^2 +{3\over 5}\omega kv_s\right)
\end{equation}
to become a quartic equation. This has four real solutions so long as
$\int {1\over v^5}{\partial f_0\over\partial p}p^2dp > 0$. In fact if
$\int {1\over v^5}{\partial f_0\over\partial p}p^2dp < 0$, the addition of higher order
terms in the expansion of the cosmic ray distribution function would dramatically
alter the character of the solutions, a situation that must be considered unphysical.
We require $\int {1\over v^5}{\partial f_0\over\partial p}p^2dp > 0$, which means
that at low momenta, $\partial f_0/\partial p > 0$, and the cosmic ray distribution
cannot be monatonically decreasing from $v=p=0$. Our use of the kappa distribution in
the preceding section may therefore be questioned. However the resonance at $p_x=-2mv_s$
places it well into the region of the distribution where the gradient is
negative, and so modifications to the low momentum behavior would have very little
effect on our result. However this observation does imply that the distribution of
particles obeying a diffusion equation ahead of the shock is unlikely to extend down to zero
momentum.
Some natural break must exist between the quasi-thermal population
gyrating around field lines and the cosmic rays diffusing in
turbulence. The forgoing also neglects the cosmic ray induced current in
the background plasma. The inclusion of such effects leads to the
modification $\omega _{pi}^2 \rightarrow\omega _{pi}^2-\omega
_{pCR}^2$, and has no effect on reactive instabilities.

\section{Discussion}
\subsection{Electron Heating or Magnetic Field Amplification?}
We have calculated the growth rate for waves that damp by heating electrons, in a
cosmic ray shock precursor using similar approximations and techniques to those employed
by \citet{Bell2004}.
Both the lower hybrid wave heating of electrons and the growth of magnetic field through
modified Alfv\'{e}n waves redistribute energy within the cosmic ray precursor. An important
question is which of
these is more effective, i.e. which grows faster? Above we derived a critical Alfv\'en
Mach number, $M_A\simeq 3n_{CR}^{\prime}/n_{CR}$, which divides the regime of magnetic
field amplification from that of lower hybrid wave growth. Following from the
treatment of the reactive instability above, we estimate $n_{CR}=\int _{p_{inj}}^{\infty}
f_{CR}4\pi p^2dp\simeq \left(6/\pi\right) n_{CR}^{\prime}v_s/v_{inj}$
where $f_{CR}$ is given by equation 1 and $p_{inj}$ is the injection momentum where
particles may begin to participate in a diffusive shock acceleration process. The
approximate result $n_{CR}^{\prime}/n_{CR}\simeq v_{inj}/2v_s$, gives
$M_A\simeq 1.5v_{inj}/v_s$ as the critical Alfv\'en Mach number.

The next step in determining the critical Alfv\'en Mach number is to find an appropriate
$v_{inj}$ for the injection of seed particles into the cosmic ray acceleration process.
\citet{Zank2006} argue that quasi-perpendicular shocks have similar injection requirements
to quasi-parallel shocks, but that highly perpendicular shocks require much higher
injection energies. In the case of nonresonant magnetic field generation, we consider
the case of a highly perpendicular shock since the generated magnetic field will be
perpendicular and much stronger than the initial magnetic field. \citet{Zank2006}
give the injection velocity as
\begin{equation}
v_{inj}=3v_s\left[{1\over\left(r-1\right)^2}+{D_{Bohm}^2\over D_{\perp}^2}\right]^{1/2}
\end{equation}
where $r$ is the shock compression ratio, and $D_{Bohm}$ and $D_{\perp}$ are
the cosmic ray diffusion coefficients in the Bohm limit and in the perpendicular
direction respectively. \citet{reville08} give $D_{Bohm}/D_{\perp}\simeq 3$
for cosmic rays where $kr_g\sim 1$, and so $v_{inj}\simeq 10 v_s$. Thus the Alfv\'en Mach
number at which lower hybrid wave growth takes over from magnetic field amplification
should be about 15, unless the magnetic field saturates at a lower value (i.e. higher
$M_A$) before this is reached.

The growth of lower hybrid waves is most efficient at a quasi-perpendicular shock, whereas
the growth of magnetic field through modified Alfv\'en waves is strongest at a quasi-parallel
shock \citep{Bell2005}. This apparent contradiction is actually easily resolved.
At an initially quasi-parallel shock, \citet{Bell2005},
\citet{reville08} and \citet{zirakashvili08} show
that a highly helical magnetic field develops.
The distortion of an initially parallel
field line is shown schematically in Figure 1, showing the evolution of the shock from
quasi-parallel to quasi-perpendicular. A similar schematic in Figure 2 shows the
evolution of an initially quasi-perpendicular shock, where magnetic field is amplified
orthogonal to the pre-existing magnetic field, but where the shock remains quasi-perpendicular.
In both cases a perpendicular field is generated, thus allowing lower hybrid wave growth and electron heating
in a region close to the shock as indicated.

Another potential problem is the cavities seen in simulations of the
growth of modified Alfv\'{e}n waves \citep[e.g.][]{Bell2005}. The
helical field from an initially quasi-parallel geometry naturally
creates a filamentary structure, dragging the thermal plasma with
it, while cosmic rays tend to accumulate in the low density
cavities. This is problematic for our mechanism that requires
spatial coincidence between cosmic rays, magnetic field and thermal
plasma. A possible solution is that the growth of lower hybrid waves
takes over from the growth of modified Alfv\'en waves, so that the
cosmic ray driven magnetic field never reaches its final saturated
state. \citet{Bell2004}, \citet{reville08} and
\citet{zirakashvili08} derive a saturation magnetic field by setting
$\gamma _{B,nonres}=0$ in equation (9) to give $\delta B\sim
jr_g/4\pi$ or $\delta B^2/8\pi\sim n_{CR}m_iv_sv_{inj}/2$. This
gives an Alfv\'en Mach number at saturation of $M_A^2\sim
n_i/10n_{CR}$, (assuming $v_{inj}\sim 10 v_s$), which for likely
parameters $n_i/n_{CR}\sim 10^3$ gives a value of $M_A$ of similar
magnitude but possibly lower than that where the electron heating is
expected to take over. Bearing in mind that we took the strongest
growth rate for magnetic field amplification to estimate where
electron heating takes over, it is quite plausible that the
amplified magnetic field never reaches saturation. Also, as the
initial shock state becomes more quasi-perpendicular, the growth
rate slows down, and the circularly polarized Alfv\'en waves become
elliptically polarized, ultimately becoming linearly polarized in
the limit of a true perpendicular shock, eliminating the growth of
such cavities.

\citet{Pelletier2006} find that the nonresonant instability of
\citet{Bell2004} and \citet{Bell2005} dominates over the more familiar resonant instability
when the shock velocity $v_s$ is greater than a few times $\epsilon _{CR}c$ where
$\epsilon _{CR}$ is the ratio of the cosmic ray energy density to the kinetic energy
density of the shock. \citet{niemiec08} simulated the cosmic
ray driven amplification of magnetic field in a parallel shock using Particle-In-Cell
simulations, which can naturally account for the backreaction of the
generated magnetic field
on the cosmic ray current. In conditions where the nonresonant mode should grow,
they find magnetic field amplification only to $\delta B\sim B$. The magnetic field
again produces filaments, but they do not find
cosmic ray accumulation
in the filament
cavities. They do not find strong growth, and argue that
saturation occurs because the incoming flow to the shock is decelerated by the
cosmic rays, reducing their relative velocity and hence the cosmic ray current.

For our electron heating model, the precise degree of magnetic field
amplification is unimportant so long as the cosmic ray diffusion
coefficient remains proportional to $1/B$. It is only necessary that
the shock be sufficiently quasi-perpendicular to allow cosmic rays
to generate lower hybrid waves. A reduced cosmic ray current does
not necessarily produce an appreciable affect on the kinetic growth
rate for lower hybrid waves. So long as the current does not vanish,
the initial effect of reducing $v_s$ in equation 6 is to bring more
cosmic rays into resonance with the lower hybrid waves. Another
estimate of the cosmic ray density necessary to heat electrons may
come from the long wavelength limit of the magnetic field
amplification, when $\gamma _B =\sqrt{{\bf k}\cdot {\bf
B}n_{CR}qv_s/n_im_i}$, for both parallel and perpendicular cases
\citep{Bell2005}. Electron heating then requires $\gamma\sim\Omega
_in_{CR}/n_i >\sqrt{n_{CR}v_s\cos\phi /n_iv_{inj}}\Omega _i$, taking
$k=\Omega _i/v_{inj}$ (probably an overestimate), yielding
$n_{CR}/n_i > \cos\phi /10$. At $\cos\phi\le\sqrt{m_e/m_i}$,
the values typical for lower hybrid wave propagation, the value
for $n_{CR}/n_i$ is low enough (0.001 - 0.01) to make electron heating
by cosmic rays plausible.

\subsection{Other Electron Heating Mechanisms}
Several other researchers have considered the generation of waves in a shock precursor
as a means of heating electrons. \citet{ohira07} and \citet{shimada00} have both
considered the model of \citet{Cargill88} in more detail, using Particle-In-Cell
codes rather than a hybrid approach. Other references
\citep{dieckmann00, mcclements01,schmitz02} focus more on the
electron injection problem for diffusive shock acceleration, rather than the thermal
electron temperature, again invoking various wave modes in a reflected ion precursor.
Our principle departure from these works has been to treat similar wave modes upstream
of the shock, but excited by cosmic rays undergoing diffusive
shock acceleration rather than by quasi-thermal ions reflected from the shock. This
allows electron heating to occur over a much more extended upstream region dictated
by the cosmic ray diffusion coefficient, $D$, rather than the ion gyroradius.  In addition,
expressing the thickness of this region as $l\sim D/v_s$ naturally results in electron
heating that is essentially independent of shock speed, as argued from observations of
Balmer-dominated shocks in \citet{GLR07}. On the other hand if the thickness of the electron
heating region is comparable to the ion gyroradius then $l\propto v_s$
and $T_e\,\propto\,v_s^2$.
This results in constant $T_e/T_i$ with $v_s$,
contrary to what is observed.

A number of other authors have investigated the role of the cross-shock potential
in heating the electrons. Inside the (quasi-perpendicular) shock ramp, the magnetic field
may ``overshoot'', i.e. increase to a value greatly in excess of its asymptotic downstream
strength before decreasing again. The electric field arising from the small charge
separation associated with this magnetic field
gradient, $E\simeq \partial/\partial x \left(B^2\right)/(8\pi e n_i)$,
can decelerate ions and accelerate
electrons. Such effects are known to be important at low Mach number shocks where a
laminar approximation holds \citep[e.g.][]{scudder86}. At higher Mach numbers, where the
shock is
turbulent, the importance of such electric fields is less clear. Electron ${\bf E}\times {\bf B}$ drift along the shock
front will result in periods of energy loss as well as energy gain by the cross-shock
potential, and hence no net heating. It has been argued \citep{gedalin07} that in certain cases the
shock front may be sufficiently thin (length scales of order $c/\omega _{pe}$) that
the electrons are effectively demagnetized. One might expect to see electron heating
{\em increase} with shock velocity (or $M_A$) once this condition becomes satisfied.
Examination of solar wind shocks suggests that such thin shocks are rare at best and certainly not ubiquitous.
We see no evidence for an increase in electron heating in SNRs up to shock velocities of 6000 km s$^{-1}$
\citep[1E 0102.2-7219;][]{hughes00}, and possibly up
to 20,000 km s$^{-1}$ \citep[SN 1993J;][]{fransson96}. At higher Mach numbers such as those
expected in gamma ray burst afterglows the convective electron gyroradius may easily reduce to
less than the electron inertial length, making the cross shock potential a candidate
heating electron heating mechanism.

\citet{Schwartz88} have made a survey of a number of solar wind shock crossings observed
{\it in situ}. They find $T_e/T_i\propto 1/M_A$ for $M_A$ greater than about 2-3. At lower $M_A$,
there is a wide scatter in $T_e/T_i$ about $T_e/T_i\sim 1$. At these slower shocks,
$T_e$ correlates very well with the change in ion velocity squared, suggesting that both
are due to the same mechanism, presumably the cross shock potential. The
switch to $T_e/T_i\propto 1/M_A$
at $M_A\sim 2-3$ is possibly due to the onset of
turbulent shock structure
at higher Mach numbers.

We can explore the conditions required for the validity of the laminar approximation
by adopting the criterion of \citet{tidman71} for the existence of a magnetosonic soliton;
\begin{equation}
M_S^2/\left(M_S^2-1\right)<M_A^2<4M_S^2\left(M_S^2-4M_S+3+2\ln M_S\right)/
\left(M_S^2-1-2\ln M_s\right)^2.
\end{equation}
The relationship between $M_A$ and $M_S$ predicted by this relation is plotted in
Figure 3.  The criterion above indicates that the laminar approximation breaks down
at slightly lower Mach numbers in the solar wind shocks than indicated by the behavior of $T_e/T_i$ in
\citet{Schwartz88}.  Magnetic field amplification by about an order of magnitude in SNR
shocks for the 400 km s$^{-1}$ shocks observed in the Cygnus Loop, where
\citet{GLR07} find complete electron-ion equilibration, would bring $M_A$ down to the
same range as indicated by \citet{Schwartz88}, possibly suggesting that the cross-shock
potential is at work for the lower velocity SNR shocks. At the higher velocity
shocks in the \citet{Schwartz88} sample, which are all perpendicular, an empirical relationship
$T_e/T_i\propto 1/M_A$ emerges.
Such a behavior can be consistent with our model if we make the assumption that
the cosmic rays accompanying solar wind shocks are
non-relativistic, suprathermal particles.
Then the
diffusion coefficients take on an extra factor $v_s/c$, assuming that the cosmic
ray velocity is proportional to the shock velocity. This extra power of the shock
velocity in the diffusion coefficient results in $T_e\propto v_s$ and $T_e/T_i\propto 1/v_s$.

\subsection{Heating versus Damping of Lower Hybrid Waves and the Width of the Precursor}

We have discussed whether growth of lower hybrid waves may compete with cosmic ray induced
magnetic field amplification, but have not yet discussed whether
this growth rate is sufficient to balance the damping rate of lower hybrid waves by electrons.
To answer this question we compare the electron heating rate from diffusive scattering off the
lower hybrid waves with the energy input into the lower hybrid waves from the cosmic ray
turbulence.
The electron heating rate per unit area of shock is $n_{e}f_R m_e\varkappa _{\|\|}tv_s/2$, where
$f_R\simeq \exp\left(-\omega^2/ 2k_{\|}^2 v_{te\|}^2\right)\simeq \exp\left(-2v_s^2m_i/v_{te}^2m_e
\right)$ is the fraction of electrons in resonance
with the lower hybrid waves (using $\omega /k\simeq 2v_s$),
$\varkappa _{\|\|}$ is the parallel electron velocity diffusion coefficient in lower hybrid
turbulence, and $t$ is the period of time spent by an electron in the turbulence. This time
$t=l/v_s$ where $l$ is the precursor depth. The electron heating is balanced by
energy input to the turbulence by cosmic rays with rate $2\gamma E_{turb}l$.
Putting $E_{turb}=\left(\delta E^2/8\pi\right)^2\omega _{pe}^2/\Omega _e^2$ and
$\varkappa _{\|\|}= q^2\delta E^2k_{\|}^2/4m_e^2k_{\perp}^2\omega $ we deduce a growth rate
$\gamma =f_R q^2B^2/16m_im_e\omega c^2 =f_R \omega /16$.
The kinetic growth rate derived earlier in Equation (7), 
in units of $\Omega_{i}$, is proportional to $n_{CR}/n_i$.  
Thus as long as this ratio is comparable to or larger than the 
fraction of electrons that are in resonance with the lower hybrid 
waves the growth rate outlined above will be sufficient to heat the
electrons.

Another constraint on $n_{CR}/n_i$ comes from Equation (9). For
magnetic field amplification, we require $n_{CR}/n_i >
k_{\|}v_A^2/v_s/\Omega _i$. Since $k_{\|}
> 1/r_g$, the gyroradius of cosmic rays at injection, $n_{CR}/n_i >
v_A^2/v_{inj}v_s\sim 1/10M_A^2\sim 10^{-3}$ for $M_A\sim 10$ and
$v_{inj}\sim 10 v_s$. Taking the maximum growth rate estimated from
Equation (9), $\gamma = M_A\Omega_in_{CR}/2n_i\sim\Omega _i/20M_A$,
where $l_i=c/\omega _{pi}=v_A/\Omega _i$ is the ion inertial length,
we estimate a characteristic length of $v_s/\gamma = 20M_A^2l_i\sim
5\times 10^{10}$~cm.  This requires a cosmic ray diffusion
coefficient of order $10^{19}$~cm$^2$~s$^{-1}$. This is considerably
smaller than the estimate by \citet{Bell2004}. Taking a
characteristic cosmic ray energy of $10^{15}$~eV, \citet{Bell2004}
finds a typical growth time for magnetic field of order 100 years.
This would yield a characteristic length scale for magnetic field
amplification of $\sim 10^{18}$~cm for a 3000 km s$^{-1}$ shock,
requiring a cosmic ray diffusion coefficient of $\sim 3\times
10^{26}$~cm$^2$~s$^{-1}$. We suspect that our simple estimate
reflects the growth rate while the shock may be considered
quasi-parallel, and that magnetic field amplification slows down
considerably once it becomes quasi-perpendicular. Therefore in
taking a characteristic cosmic ray energy of $10^{15}$~eV,
\citet{Bell2004} is taking the lowest energy cosmic rays for which
the shock may be considered quasi-parallel, and this result may be
considered more realistic.

Further, in \citet{GLR07} we argued that the depth of cosmic ray precursor
over which electron heating occurs could not be larger than $\sim 10^8v_s/n_e$ cm, otherwise
neutral hydrogen would not survive to encounter the shock front. We suggest here
that lower hybrid
waves accelerate the small fraction of electrons that happen to be in resonance, and that
these accelerated electrons communicate their energy to the rest of the thermal population
by Coulomb collisions, with characteristic time scale
$10^{10}\left(T/10^8 {\rm K}\right)^{3/2}/n_e$  s.
Equating this to $10^8/n_e$ s yields a maximum temperature of
$T\sim 10^8\times\left(10^{-2}\right)^{2/3}\simeq 5\times 10^6$ K. This is very close
to the temperature found in \citet{GLR07}, 0.3 keV, or $3.5\times 10^6$ K. Put another
way, the temperature found in \citet{GLR07} is consistent with electron heating such
that neutral hydrogen can survive to encounter the shock front proper. However
cosmic ray precursors at the small end of the range considered above ($\sim 10^{11}$~cm)
would not allow any significant electron collisional equilibration to occur.
Allowing for compressional heating of the electrons as they go through the shock,
a precursor electron temperature of order $10^6$~K requires a precursor length of
$\sim 10^7v_s/n_e\sim 10^{15}\left(v_s/1000 {\rm ~km~s}^{-1}\right)$~cm, or a
minimum cosmic ray diffusion coefficient of $D\sim 10^{23}
\left(v_s/1000 {\rm ~km~s}^{-1}\right)^2$~cm$^2$s$^{-1}$. 

The electric field in the lower hybrid waves will be given by the limit derived
by \citet{karney78},
\begin{equation}
\delta E=B\left(\Omega _i\over\omega\right)^{1/3}{\omega\over 4k_{\perp}c}=
B\left(\Omega _i\over\omega\right)^{1/3}{v_s\over 2k_{\perp}c}.
\end{equation}
This is the maximum electric field before ion trapping and heating occurs. \citet{laming07}
and references cited therein demonstrate that when $\omega /\sqrt{2}k_{\|}v_{te} <<
\omega/\sqrt{2}k_{\perp}v_{ti}$, ions are heated more effectively than electrons above this
threshold. In our case, the ions that are heated will be the lower energy part of the
suprathermal ion distribution reflected from the shock, i.e. those below the injection
threshold for diffusive shock acceleration in Equation (1) or any of its modifications
subsequent to the treatment of the reactive lower hybrid wave instability in section 2.2.
With the wave electric field given by Equation (27), the electron momentum diffusion
coefficient in lower hybrid turbulence varies as $v_s^2$, yielding a constant degree
of heating with shock velocity if the time spent in the turbulence varies as $1/v_s^2$,
which would be the case if the cosmic rays are obeying a diffusion law.

\section{Summary}
We have considered in more detail the speculation of \citet{GLR07} that lower hybrid
waves generated in a cosmic ray precursor could be responsible for the electron
heating at collisionless shocks in supernova remnants.
We find that there do exist growing modes for the resonant or kinetic case, and that
the growth rate in this case may be sufficient both to survive the damping by electrons
and to compete with magnetic field amplification by modified Alfv\'en waves.
Below a certain Alfv\'en Mach number (roughly estimated to be $\sim 15$)
the lower hybrid wave growth rate exceeds that of the modified Alfv\'{e}n waves.
The modified Alfv\'en wave generation exists for all magnetic field orientations with
respect to the shock, but is most effective for quasi-parallel case and always generates new perpendicular field.
Lower hybrid waves, on the other hand, require quasi-perpendicular field geometry in order to grow.
Thus a schematic picture emerges in which far ahead
of the high Mach number shock, modified Alfv\'en waves generate perpendicular field, reducing the effective Mach
number closer to the shock front and thus allowing lower hybrid wave growth to occur in a short
region before the shock and heat the resonant electrons.
A critical Alfv\'en Mach number around 15 suggests magnetic field amplification by about
an order of magnitude, similar to what a comparison of the surveys of \citet{GLR07} and
\citet{Schwartz88} would suggest, taking in both cases the shock velocity where \tetp $\sim$1
starts to break down as that where the laminar shock approximation ceases to hold.

We have concentrated on the generation of lower hybrid waves, since
for these the group velocity can be equal to the shock velocity itself, meaning that
the waves can stay in contact with the shock for long time intervals and in principle
grow to large amplitudes. However other wave modes that heat electrons are certainly
possible, and these, such as the Landau damping of kinetic Alfv\'en waves
\citep[e.g.][]{vinas00}, do not
require perpendicular shocks as lower hybrid waves do. In fact, \citet{bykov} studied the
generic case of heating by turbulent modes in the shock precursor and did identify an area
of parameter space for which a near inverse-square relationship between \tetp and shock
velocity could be accommodated.
Our model requires that cosmic ray ions be essentially ubiquitous at 
SNR shocks, with number densities estimated by various means in section 3.
In a wider context, the idea that cosmic rays are responsible for electron heating
at fast shocks reinforces the idea that cosmic rays are an intrinsic component of the
collisionless shock phenomenon.

\acknowledgments
J.M.L. and C.E.R. have been supported by NASA contract NNH06AD66I (LTSA Program)
and by basic research funds of the Office of Naval Research. P.G. acknowledges support
from NASA contract NAS8-03060. We also appreciate the continuing advice and
encouragement of Dr. Jill Dahlburg.

\appendix

\section{Cosmic Ray Diffusion Coefficients}
The parallel spatial cosmic ray diffusion coefficient is most easily obtained from
its relation to the pitch angle scattering diffusion coefficient in momentum space.
The diffusion coefficient in momentum space is expressed most generally as
\citet{melrose86},
\begin{equation}
D_{\lambda\mu} = \sum_{s=-\infty}^{\infty} \int
\frac{8\pi ^2q^{2}}{\hbar}
\frac{R_{M}({\bf k})}{\omega_{M}({\bf k})}
\vert {\bf e} \cdot {\bf v}({\bf k},{\bf p}, s)\vert ^{2}
\delta (\omega_{M} - s\Omega - k_{\|}v_{\|})\triangle\lambda\triangle\mu
N_{\mu}({\bf k})\frac{d^{3}k}{(2\pi)^{3}}
\end{equation}
where $N_{\mu}$ is the number density of wave quanta, $R_M$ is the ratio of electric
energy to total energy in the wave, such that $R_M\int N_{\mu}\hbar\omega _Md^3k/\left(2\pi\right)^3
=\delta E^2/8\pi $, ${\bf e}$ is the wave polarization vector and ${\bf v}$ is the cosmic ray
velocity. For pitch angle scattering by parallel propagating Alfv\'{e}n waves,
$\lambda=\alpha$, and so
\begin{equation}
\triangle\lambda =
\hbar\left(\frac{s\Omega}{v_{\bot}}\frac{\partial}{\partial
p_{\bot}} + k_{\|}\frac{\partial}{\partial
p_{\|}}\right)\lambda =-{\hbar k_{\|}\over
p\sin\alpha} .
\end{equation}
With $\omega_{M} = k_{\|}v_A$, $R_{M} = (v_{\|}^{2})/(2c^{2})$, and
${\bf e}\cdot{\bf v}= v_{\bot}/2$,
\begin{eqnarray}
\nonumber D_{\alpha\alpha} & = & \int \frac{8 \pi ^2
q^{2}}{\hbar\omega}\frac{v_A^{2}}{2c^{2}}
\frac{v^{2}\sin^{2}\alpha}{4} \delta (\omega_{M} - s\Omega -
k_{\|}v_{\|}) \frac{\hbar^{2}k_{\|}^{2}}{p^{2}\sin^{2}\alpha}
\frac{U_{M}({\bf k})}{\hbar \omega}
\frac{d^{3}k}{(2\pi)^{3}} \\
& = & \frac{\pi ^2q^{2}v} {p^{2}c^2 \cos\alpha}
\frac{U_{M}(k_{\|}=\Omega/v_{\|})}{2\pi}
\end{eqnarray}
where we have put, $s=1$ and taken $\omega _M << \Omega$.

We now express $D_{\|}$ in terms of $D_{\alpha\alpha}$ by writing
\begin{equation}
f(p,\alpha)=f_{0}(p)+f_{1}(p)\cos\alpha+\onehalf f_{2}(p)\cos^{2}\alpha + ...
\end{equation}
and substituting into the diffusion equation
\begin{equation}
\frac{\partial f}{\partial t} + v_{z}\frac{\partial f}{\partial z} =
\frac{1}{\sin \alpha} \frac{\partial}{\partial \alpha}
\left[\sin\alpha D_{\alpha\alpha} \frac{\partial f}{\partial \alpha} \right].
\end{equation}
%
Upon integrating the result over $\cos\alpha$ we obtain, with $v_z=v\cos\alpha$,
\begin{equation}
\frac{\partial f_{0}}{\partial t} +\frac{1}{3}\frac{\partial f_{2}}{\partial t}
+ \frac{v}{3}\frac{\partial f_{1}}{\partial z} = 0.
\end{equation}
Multiplying each side by $\cos \alpha$ and then integrating over $\cos\alpha$ yields
\begin{equation}
\frac{2}{3}\frac{\partial f_{1}}{\partial t} +
\frac{2v}{3}\frac{\partial f_{0}}{\partial z}+
\frac{v}{5}\frac{\partial f_{2}}{\partial z} =- \int_{-1}^{1}\cos\alpha\sin^{2}\alpha Df_{2}d(\cos\alpha).
\end{equation}
With $f_0 >> f_1 >> f_2$,
$f_{2} \simeq -
\frac{2v}{3}\frac{\partial f_{0}}{\partial z}
\left[ \int_{-1}^{1}\cos\alpha\sin^{2}\alpha Dd(\cos\alpha) \right]^{-1}$ which when substituted into
equation A6 allows the identification
\begin{equation}
D_{\|}={2v^2\over 9}\left[\int _{-1}^{1}\cos\alpha\sin ^2\alpha D_{\alpha\alpha}d\left(\cos\alpha\right)
\right]^{-1}.
\end{equation}
With equation A3,
\begin{equation}
D_{\|}={p^2c^2v\over 3\pi q^2 U_M\left(k_{\|}=\Omega
/v_{\perp}\right)}.
\end{equation}
This is a factor of $2\pi$ larger than the equivalent expression
given by \citet{blandford87}, due to a different definition of
$U_M$. Where $U_M\propto k_{\|}^{-\beta }$, $D_{\|}\propto
p^{2-\beta}$, which evaluates to $D_{\|}\propto vp^{1/3}$ or
$D_{\|}\propto vp^{1/2}$ for Kolmogorov or Kraichnan turbulence
respectively. If $v\sim c$, the dependence of $D_{\|}$ on $p$ can
usually be neglected.

The perpendicular spatial cosmic ray diffusion coefficient has been
given in terms of $D_{\|}$ by various authors. Based on numerical
experiments, \citet{Marcowith2006} give $D_{\perp}=\eta
^{2+\epsilon}D_{\|}$ where $\eta =\delta B^2/\left(\delta
B^2+\left<B\right>^2\right)$ and the cosmic ray distribution
function $f\left(p\right)\propto p^{-4-\epsilon}$.
\citet{Shalchi2007} and \citet{Zank2006} give
$D_{\perp}\propto\left(\delta B^2/B_0^2
\right)^{2/3}D_{\|}^{1/3}\left(l_{2D}v\right)^{2/3}$ from analytic
considerations, where $l_{2D}$ is the 2D bendover length scale, the
inverse of the wavenumber where the inertial range onsets, and
consequently has even smaller dependence on the cosmic ray momentum
than the parallel diffusion coefficient for relativistic cosmic rays, and has
the same dependence in the nonrelativistic case.

\section{Growth Rate for an Electromagnetic Instability}
For completeness, we give here a treatment of the growth rate due to
cosmic rays of electromagnetic waves with frequency in the lower
hybrid range, and show that it is significantly smaller than either
the electrostatic instability, or the growth of modified Alfv\'en
waves. It is relatively easy to show that the reactive instability
of \citet{Bell2004} has higher thresholds and lower growth rates as
the frequency of the electromagnetic wave increases first above the
proton gyrofrequency and then above the electron gyrofrequency. Here
we concentrate on the kinetic instability that might generate
electromagnetic waves in the lower hybrid range, whistlers, adapting
the expressions in \citet{Bell2004} and \citet{Achterberg83};
\begin{eqnarray}
\omega^{2}(K^{T} -1) & =  & \frac{\Omega_{i}c^{2}}{\omega v_{A}^{2}}
\left\{ \tilde{\omega}_{i}^{2}
\mp \frac{k^{2}v_{ti}^{2}}{\Omega_{i}}\tilde{\omega}_{i}
\mp \frac{\Omega_{i}}{n_{e}}k\left( \frac{J_{CR}}{q} - \frac{\omega}{k}N_{CR}
\right) \right\}
 \\
& + & 4\pi q^{2} \int \frac{v_{\bot}/2}{\omega - k_{\|}v_{\|}}
\left\{ (\omega-k_{\|}v_{\|} ) \frac{\partial f}{\partial p_{\bot}}
+ k_{\|}v_{\bot} \frac{\partial f}{\partial p_{\|}} \right\} 2\pi
p_{\bot}dp_{\bot} dp_{\|} = c^{2}k^{2}-\omega ^2. \nonumber
\end{eqnarray}
Evaluating the two cosmic ray terms,
\begin{equation}
\int\int
v_{\bot}\frac{\partial f}{\partial p_{\bot}}2\pi p_{\bot}dp_{\bot}dp_{\|}
=\int_{-\infty}^{\infty}\left\{\left[2\pi p_{\bot}v_{\bot}f \right]_{0}^{\infty}
-\int \frac{4\pi p_{\bot}}{\gamma m}f dp_{\bot}
\right\} dp_{\|} = -\frac{2}{\gamma m}n_{CR}
\end{equation}
and
\begin{eqnarray}
\int\int \frac{v_{\bot}^{2}}{(\omega-k_{\|}v_{\|})}
\frac{\partial f}{\partial p_{\|}}2\pi p_{\bot}dp_{\bot}dp_{\|}
& =  & \int \int
\frac{-p_{\bot}^{2}/(\gamma^{2}m^{2})}{(\omega-k_{\|}v_{\|})}
\frac{(p_{\|}-mv_{s})/p_{t}^{2}}
{\left[1+ ((p_{\|}-mv_{s})^{2}+ p_{\bot}^{2})/(2\kappa p_{t}^{2})\right]^{\kappa+1}}
2\pi p_{\bot}dp_{\bot}dp_{\|}
\nonumber \\
& = &  \int \int
\frac{p_{\bot}(p_{\|}-mv_{s})/(\gamma^{2}m^{2})}{(\omega-k_{\|}v_{\|})}
\frac{\partial f}{\partial p_{\bot}}2\pi p_{\bot}dp_{\bot}dp_{\|}
\end{eqnarray}
explicitly assuming a $\kappa$ distribution for CR in Equation (B3).
After some algebraic manipulation, an integration by parts,
rewriting $f$ as a $\kappa$ distribution and making the
substitution $p_{\bot}^{2} = P, dP = 2p_{\bot}dp_{\bot}$ (B3)
can be written as
\begin{equation}
- \int \frac{4\pi(p_{\|}-mv_{s})}{\omega-k_{\|}v_{\|}}
\int \left(\frac{2\kappa p_{t}^{2}}
          {2\kappa p_{t}^{2}+(p_{\|}-mv_{s})^{2}+ P}\right)^{\kappa}
\frac{m^{2}c^{4}+p_{\|}^{2}c^{2}}{(m^{2}c^{2}+p_{\|}^{2}+P)^{2}}
dPdp_{\|}.
\end{equation}
The integral over $dP$ can be evaluated using a hypergeometic
function (Gradshteyn \& Ryzhik 1965; 3.197.1) to give
\begin{eqnarray}
 & - \int \frac{4\pi(p_{\|}-mv_{s})}{\omega-k_{\|}v_{\|}}
\frac{(2\kappa p_{t}^{2})^{\kappa}}{(m^{2}+p_{\|}^{2}/c^{2})}
\frac{{\rm B}(1,1+\kappa)}
          {(2\kappa p_{t}^{2}+(p_{\|}-mv_{s})^{2})^{\kappa -1}}
\nonumber \\
 \times &
 _{2}{\rm F}_{1}\left(2,1;2+\kappa;1-\frac{2\kappa p_{t}^{2}+(p_{\|}-mv_{s})^{2}}{m^{2}c^{2}+p_{\|}^{2}}\right)
dp_{\|}
\end{eqnarray}
where ${\rm B}(1,1+\kappa )$ is the beta function. Considering only
the kinetic case, using the Landau prescription for this integral
with the pole at $\omega -k_{\|}v_{\|}$, the imaginary part (i.e.
the portion relevant for the growth rate) is
\begin{eqnarray}
{\rm Im}\left(
\int\int \frac{v_{\bot}^{2}}{(\omega-k_{\|}v_{\|})}
\frac{\partial f}{\partial p_{\|}}2\pi p_{\bot}dp_{\bot}dp_{\|} \right)
& = &
4i \pi^{2} \left(\frac{\gamma m \omega}{k_{\|}} - mv_{s} \right)
\frac{(2\kappa p_{t}^{2})^{\kappa}}
      {m^{2}+ \gamma^{2}m^{2}\omega^{2}/(k_{\|}^{2}c^{2})}
\frac{m}{k_{\|}}
\nonumber \\
& \times &
\frac{{\rm B}(1,1+\kappa)}
      {(2\kappa p_{t}^{2}+(\gamma m \omega/k_{\|}-mv_{s})^{2})^{\kappa -1}}
\nonumber \\
&  \times &
 _{2}{\rm F}_{1}\left(2,1;2+\kappa;
1-\frac{2\kappa p_{t}^{2}+(\gamma m \omega/k_{\|}-mv_{s})^{2}}
{m^{2}c^{2}+\gamma^{2}m^{2}\omega^{2}/k_{\|}^{2}}\right)
.
\end{eqnarray}
Setting $\omega\rightarrow \omega + i\gamma_{g}$ in the dispersion
relation and taking only the imaginary parts we get
\begin{eqnarray}
0 = \Omega_{i}\gamma_{g}+2\omega\gamma _g \pm
\frac{\Omega_{i}}{\omega^{2}}\gamma_{g}\frac{kJ_{CR}}{n_{e}q}
   (2\pi)^{3} q^{2} k_{\|}\left(\gamma\omega m/k_{\|}- mv_{s}\right)
\frac{(2\kappa p_{t}^{2})^{\kappa}}{(2\kappa p_{t}^{2}+m^{2}v_{s}^{2})^{\kappa -1}}
{\rm B}(1,1+\kappa)
\frac{v_{A}^{2}}{c^{2}} \frac{1}{m^{2}} \frac{m}{k_{\|}}
& & \nonumber \\
\times
 _{2}{\rm F}_{1}\left(2,1;2+\kappa;
1-\frac{2\kappa p_{t}^{2}+m^{2}v_{s}^{2}}
{m^{2}c^{2}}\right)
\frac{n_{CR}}{4\sqrt{2}(\pi \kappa)^{3/2}p_{t}^{3}}
\frac{(2\kappa-3)\Gamma(\kappa)}{\Gamma(\kappa -1/2)} & &
\end{eqnarray}
assuming $\gamma\omega/k_{\|} \ll v_{s}$ and including the normalization of
$f$ and the factor of $k_{\|}/2$ that were omitted during the evaluation of
the integral.
From this we have
\begin{equation}
\gamma_{g} \simeq -\left(\frac{\pi}{2}\right)^{1/2}
\frac{n_{CR}}{n_{i}} {\Omega_{i}^2\over\omega }
\frac{\left(\gamma\omega m/k_{\|}- mv_{s}\right)}{p_{t}}
\frac{2\kappa-3}{\kappa^{1/2}(\kappa +1)}
\frac{\Gamma(\kappa)}{\Gamma(\kappa -1/2)} \times _{2}{\rm
F}_{1}(2,1;2+\kappa;1) .
\end{equation}
Electromagnetic waves in the lower hybrid frequency range are
parallel propagating whistlers, with
\begin{equation}
{k^2c^2\over\omega ^2}\simeq {\omega _{pe}^2\over\omega\left(\Omega
_e-\omega\right)}
\end{equation}
and
\begin{equation}
{\partial\omega\over\partial k}\simeq {2\omega /k\over 1+k^2c^2/\omega _{pe}^2}.
\end{equation}
We assume $\partial\omega /\partial k\propto\partial\omega /\partial k_{\|}\sim v_s$
and hence $\omega /k_{\|}\sim v_s/2$ for $k<< c/\omega _{pe}$. Thus for $\kappa =2$ and
\begin{equation}
\gamma_{g} \simeq \frac{4n_{CR}}{3n_{i}}
{\Omega_{i}^2\over\omega}\left(1-\gamma /2\right).
\end{equation}
This is significantly smaller than the growth of lower hybrid waves, which is
of order $\Omega _in_{CR}/n_i$.  Further, since
whistlers carry energy along magnetic field lines, like Alfv\'en waves, only for
specific shock obliquities will the energy of the waves stay in contact with the shock
and allow large wave intensities to build up. Electromagnetic waves with frequency
above the electron gyrofrequency (O and X modes) have phase velocities greater than
$c$, and so cannot be excited by kinetic instabilities.

\clearpage
\begin{figure}
\plotone{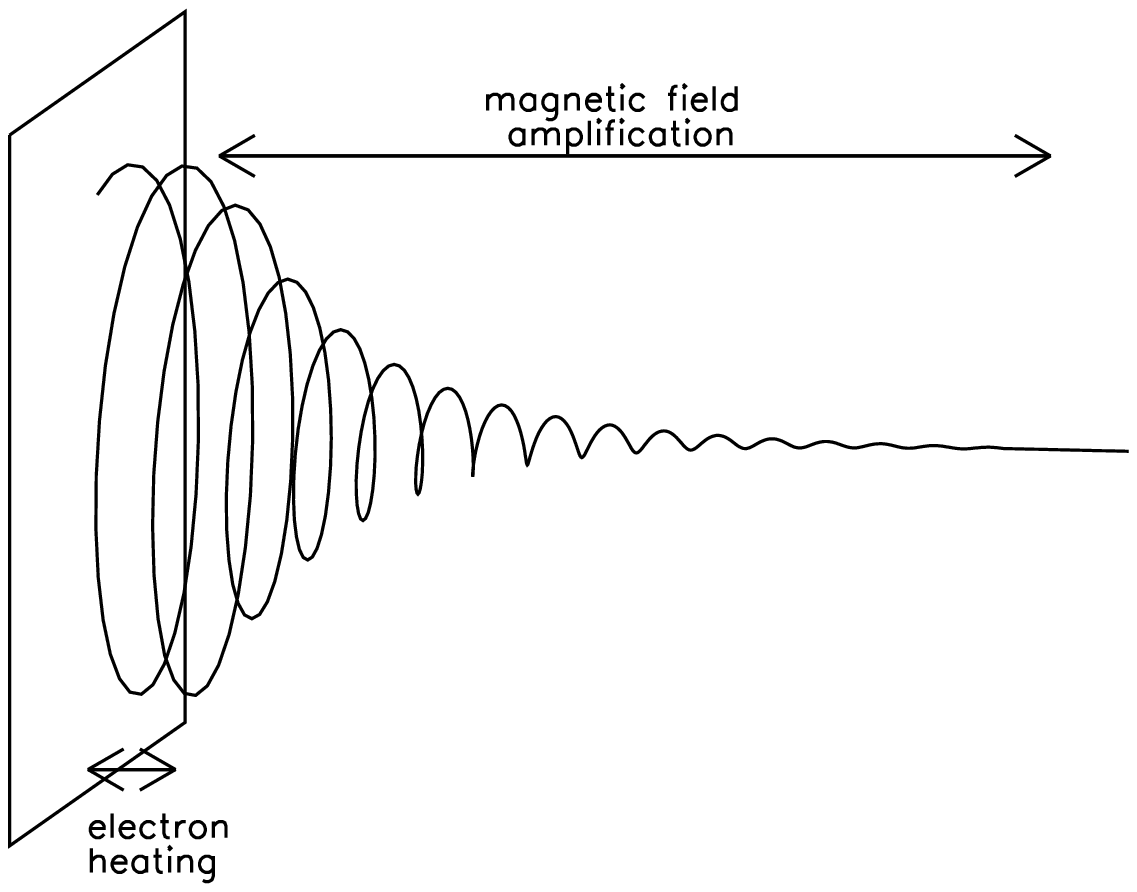} \figcaption[f1.eps]{Schematic illustrations of the
amplification of magnetic field by the non-resonant modified Alfv\'{e}n waves in the
shock precursor in the parallel orientation of the
ambient field with respect to the shock normal. The evolution of a single field line
in an exponential purely growing mode is shown.
As the field is amplified, the shock becomes quasi-perpendicular and
the effective $M_{A}$ decreases, eventually to the point where
lower hybrid wave growth takes over, allowing a short region of electron heating.
\label{fig1}}
\end{figure}

\clearpage
\begin{figure}
\plotone{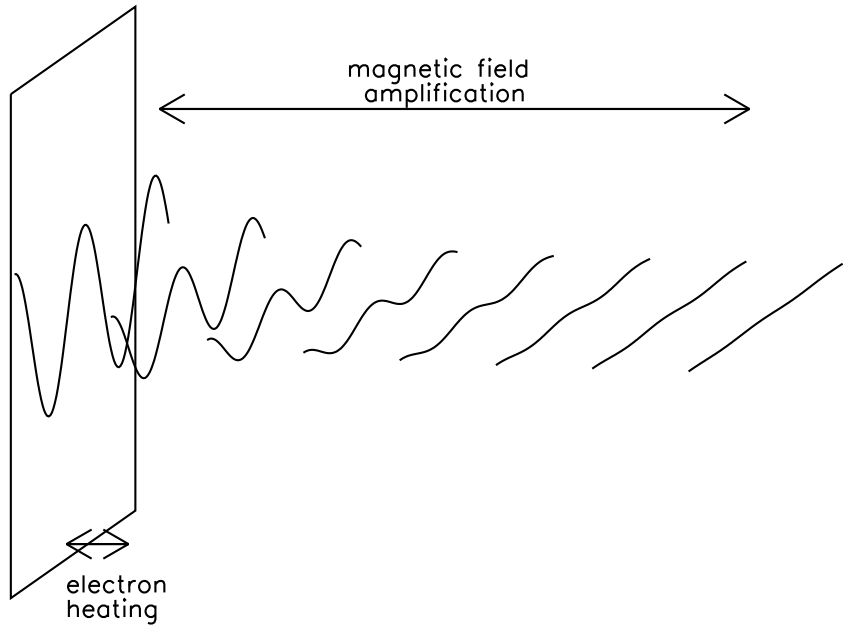} \figcaption[f2.eps]{Same as figure 1 but for an initially
perpendicular shock. The evolution of a purely growing mode is illustrated.
The magnetic field amplification is less strong than in the quasi-parallel case, and the shock geometry
remains quasi-perpendicular.
\label{fig2}}
\end{figure}

\clearpage
\begin{figure}
\plotone{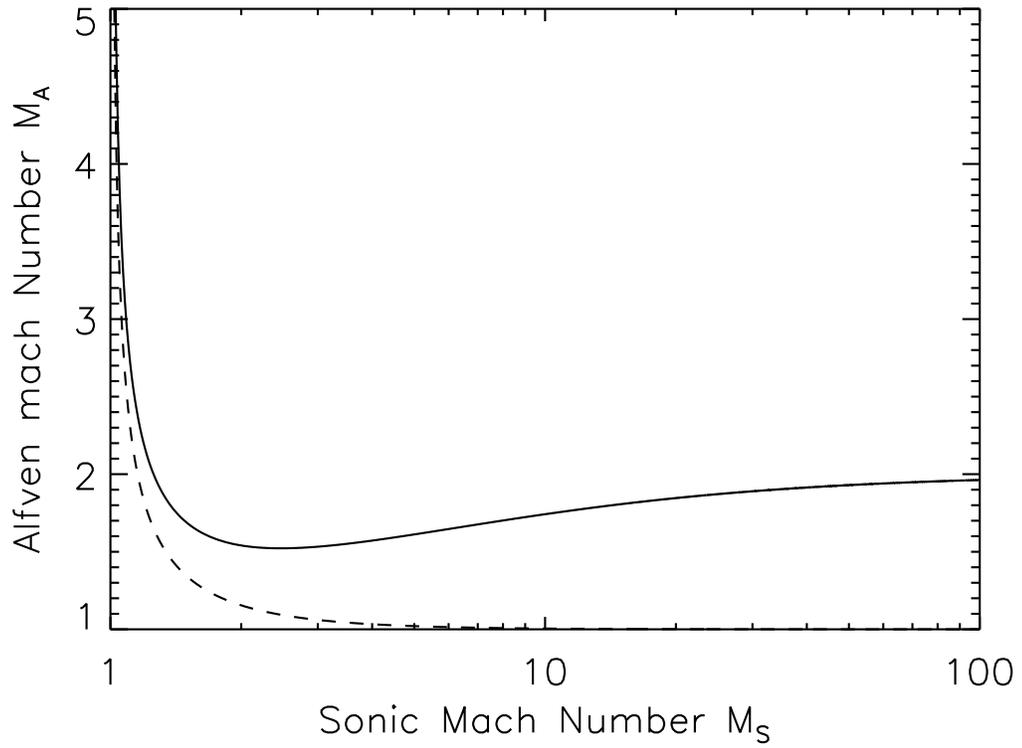} \figcaption[f3.eps]{Allowed range of $M_A$ as a function of
$M_S$ for the existence of a magnetosonic soliton, from Tidman \& Krall (1971). The upper
limit is given by the solid line, the lower limit by the dashed line.
\label{fig3}}
\end{figure}

\end{document}